\documentclass[preprint,showpacs,preprintnumbers,amsmath,amssymb,superscriptaddress]{article}
\usepackage[dvipdfmx]{graphicx}
\usepackage{dcolumn}
\usepackage{bm}
\usepackage{braket}
\usepackage{amsmath}
\usepackage{subfigure}
\usepackage{color}
\usepackage{authblk}
\usepackage{indentfirst}

\begin{document}
\title{Probing Space-time Distortion with Laser Wake Field Acceleration and X-ray Free Electron Lasers }
\author[1]{M. Yano\thanks{Corresponding author. Email address: myano@ef.eie.eng.osaka-u.ac.jp(M. Yano)}}

\affil[1]{Graduate School of Engineering, Osaka University, Yamada-oka 2-1, Suita, Osaka 565-0871, Japan}
\author[2]{A. Zhidkov}
\affil[2]{Photon Pioneers Center, Osaka University, 2-1, Yamadaoka, Suita, Osaka 565-0871, Japan}
\author[1,2]{R. Kodama}
\date{}
\maketitle

\begin{abstract}

Petawatt class femtosecond lasers and x-ray free electron lasers (XFEL) open up a new page in research fields related to space and vacuum physics. One of fundamental principles can be explored by these new instruments is the equivalence principle, saying that gravitation and acceleration should be treated equivalently. If it is true this must lead to the appearance of the Unruh effect analogous to Hawking's black hole radiation. It says that a detector moving with a constant acceleration $w$ sees a boson's thermal bath with its temperature $T_w=\hbar w/2\pi k_B c$. Practical detection of the Unruh effect requires extremely strong acceleration and fast probing sources. Here we demonstrate that x-rays scattering from highly accelerated electrons can be used for such detection. We present two, feasible for the Unruh effect, designs for highly accelerated systems produced in underdense plasma irradiated by high intensity lasers pulses. The first is Thomson scattering of a XFEL pulse from plasma wave excited by an intense laser pulse. Properly chosen observation angles enable us to distinguish the Unruh effect from the normal Doppler shift with a reasonable photon number. The second is a system consisting of electron bunches accelerated by a laser wake-field, as sources and as detectors, which move in a constantly accelerating reference frame (Rindler space) and are probed by x-ray free electron laser pulses. The numbers of photons is shown to be enough to observe {reproducible} results.\\

\end{abstract}

\newpage
		\section{Introduction}

 The existence of the Unruh radiation remains a fundamental question in quantum electrodynamics\cite{Unruh1}. Whether kinematic acceleration, in some sense, is equivalent to the gravitational field with the consequences of this principle and feasibility of Rindler space\cite{Rindler} have yet to be verified experimentally. According to Unruh\cite{Unruh1} in a reference frame moving with a constant acceleration $w$, a detector should experience surrounding vacuum as a boson’s heat bath with its temperature $T_w=\hbar w/2\pi k_B c$\cite{Unruh1}. At the first sight the Unruh effect has no practical meaning: to get a boson gas with the temperature, for example, of the Relic radiation $T_R=2.7 \ \ K$\cite{Sunyaev} the acceleration must exceed $w \approx 10^{20} g$, where $g$ is the Earth's gravitational acceleration. However, studying the highly accelerated system in the reality may allow us to understand better hidden properties of vacuum such as meaning of vacuum states, radiation transition times, and so on. Exploring the Unruh hypotheses is one of the important ways to probe the vacuum.

Since the Unruh theory has yet to be verified there are several standpoints on it: positive\cite{Bell} and negative\cite{Narozhnii}. The positive standpoints state that the Unruh effect should exist because the temperature predicted by the theory coincides with that from the Hawking’s theory of radiation of black holes\cite{Hawking}. The negative standpoints say that the Rindler space is not congruent physically with Minkowski space and the Unruh effect should not exist\cite{Narozhnii}. Indeed, a non-inertial reference system should be carefully checked on its feasibility to physical processes. For example a reference frame rotating with frequency $\Omega$ is not feasible for distance $R>c/\Omega$\cite{Landau1}. A reference frame moving with constant acceleration $w$ has also some paradoxes rising the question about its feasibility: (i) an electron moving with constant acceleration radiates while its radiation friction force is zero, the energy conservation requires existence of beginning and an end of the acceleration process\cite{Ginzburg}, (ii) apparently there is no conservative system for several particles in a reference frame moves with constant acceleration $w$; (iii) it feels as if there is no fixed photon mode $({\bf k}, \omega)$ in this reference frame due to Doppler upshift or downshift. Nevertheless the answers on the feasibility can be found only experimentally.

 Radiation in a highly accelerated system has another fundamental question which appears recently in the quantum approach for the radiation damping force during interaction of energetic electrons with high power laser pulses\cite{Ritus}. For example, Compton scattering in strong laser fields runs mainly as $n\omega_L+e^- > \omega_{x-ray}+e^-$ where n is the number of a harmonic, $\omega_L$ is the laser frequency and $\omega_{x-ray}$ is the frequency of a scattered photon. This scattering has been considered in the classical approach\cite{Ritus} and in the relativistic quantum approaches\cite{Esarey1} with one of results as $n \approx a_0$ if $a_0 \gg 1${, where $a_0$ is normalized vector potential.} Unfortunately, so far the QED approach is not available for the problem: the laser field is always considered as the classical field. Therefore, the time of Compton scattering is not known. The duration of scattering cannot be found in the frame of present approaches. This question is critical in numerical investigations of plasma interaction with high power laser pulses. Exploring of Compton scattering in a reference frame moving with constant acceleration could provide us with the answer if the detector velocities in absorption time and emission time were essentially different.
 
 Here we propose original designs for systems consisting of highly accelerated electrons which move in wake field of PW-class laser pulses and can be considered as objects in a constantly accelerating reference frame (Rindler space). Femtosecond XFEL pulses are proposed to probe such accelerated systems at extremely high acceleration  $w \approx 2 \pi c^2 a_0/\lambda$ at a high temporal resolution.
 
 Rindler space\cite{Rindler} is a natural reference frame to investigate a system moving with constant acceleration $w$ for example in $x$ direction. Coordinates in this reference frame link to those in Minkowski space as $ ct=(c^2/w+x^\prime )  \sinh⁡(wt^\prime/c)$, $ x =(c^2/w+x^\prime )  \cosh⁡(wt^\prime/c)$ with the interval $ds^2=(1+wx^\prime/c^2 )^2 d(ct^\prime)^2-({dx^\prime}^2+dy^2+dz^2) = \rho^2 d\eta^2 - (d\rho^2 + dy^2 + dz^2)$ [$g^{00}  = \rho^{-2}, g^{ii} = -1, \sqrt{-g} = \rho$]. Unruh used a different form for the metric in \cite{Unruh1}: $ds^2 = \rho d\eta^2 - (d\rho^2/\rho + dy^2 + dz^2)$, $g^{00} = 1/\rho$, $g^{11} = -\rho$, $\sqrt{-g} = 1$. Klein-Gordon equation in a curved space has a following form: $\sqrt{-1/g} \ \ \partial/{\partial x^\mu}  \left[g^{\mu\nu} \sqrt{-g}  \ \ \partial / \partial x^\nu\right]\psi=(m^2 c^2/\hbar^2)  \psi$. [For massless particles, this equation does not contain the Plank constant.] For example, in Rindler coordinates Klein-Gordon equation for a massless particle coincides with an equation for potentials of electro-magnetic field and has a following form: $[\partial^2 /\partial\eta^2 - \rho (\partial / \partial \rho) \rho (\partial / \partial \rho) - \partial^2 / \partial y^2 - \partial^2 / \partial z^2 ] \psi = 0 $. One can easily check that a plane wave solution of Klein-Gordon equation in Minkowski space  $\psi \sim e^{i\omega (t - x/c)} = e^{i \rho[\cosh⁡(\eta)-\sinh⁡(\eta)]}$ obeys Klein-Gordon equation in Rindler space. However, in Rindler space there are other solutions of KG equation in the form of Fulling modes: $\psi_{FR} \sim g(\rho)e^{i(\varpi \eta - k_y y-k_z z)}$ where $g(\rho)$ obeys following equation\cite{Belinskii}
 	\begin{equation}
	\left[\rho \frac{\partial}{\partial \rho} \rho \frac{\partial}{\partial \rho} + \varpi^2 -{k_y}^2 - {k_z}^2 - \tilde{m}^2 \right] g = 0 ,
	\end{equation}

 These modes can be used for the second quantization in Rindler space similar to that with the plane waves in Minkowski space. It is the reasonable assumption that a detector moving with a constant acceleration should see vacuum as a set of Fulling modes. This is the basis of the Unruh effect. The same results occur while considering Maxwell's equations and Dirac equations.

According to Unruh for {an observer with constant acceleration}, the vacuum state is changed and the detector observes the boson distribution of Rindler particles $\hat{\rho} = \rho_0 e^{-2\pi \hat{K}} \equiv \rho_0  e^{- \hat{K}_w/k_B T_w } $in the vacuum. Where $\rho_0 = \prod_j[(1-e^{-2 \pi \varpi_j } )]$, $\hat{K}_w = \hbar w \hat{K} ̂$, $ \varpi_j$ is frequency of particle and $\hat{K}$ is the energy of massless scalar particles. The temperature is called the Unruh temperature given by $T_w=\hbar w/2\pi k_B c$. There are papers exploring the effect\cite{Crispino}. Unruh and Wald have shown that the particle bathed in boson's thermal bath is excited absorbing Rindler particles and {really emits photons}\cite{Unruh2}. This is called the Unruh radiation. Letaw and Pfautsch considered the excitation of the particle moving circularly with constant velocity\cite{Letaw}. Chen and Tajima have shown that the Unruh radiation from electron accelerated with extremely high gradient by the static wave produced by intense laser could be observed\cite{Chen}. Chen and Mourou have recently proposed an interesting application for relativistic plasma mirror stopped abruptly by impinging intense x-ray pulses\cite{Chen2}. Also, B. J. B. Crowley { et al.} showed that the spectral broadening due to acceleration of X-ray Thomson scattering from electrons in plasma accelerated by intense laser could detect the Unruh radiation\cite{Crowley}. According to Zel'dovich\cite{Zel'dovich} an electron moving with acceleration $w$ crosses a photon bath with its energy density $W \sim T^4 \sim w^4$. This electron should scatter such photons with the known Thomson cross section $\sigma = 10 \alpha^2/m^2$ [$\alpha$ is the fine-structure constant] with total power $I \sim 10 \alpha^2 w^4/m^2$ which may exceed the classical radiation of the electron $I \sim 2 \alpha w^2/3$ {when acceleration $w$ exceeds $\sqrt{m^2/15 \alpha}$}. Besides, the electron may annihilate with a positron from the electron-positron bath emitting a characteristic radiation. Ginzburg and Frolov\cite{Ginzburg}  have found an analogy between the Unruh effect and the anomalous Doppler effect.
 
 Besides, there are problems related to the Unruh radiation. First of all, the Unruh effect of all particles except photons is purely quantum effect. However, there is no definition of acceleration $w$ in quantum mechanics as discussed by Narozhnii { et al.}\cite{Narozhnii} According to Narozhnii { et al.}, there is no conformity of Rindler space-time to Minkowski space-time\cite{Narozhnii, Belinskii}. In a certain boundary condition, the Unruh effect therefore cannot exist\cite{Narozhnii}. Nikishov and Ritus have shown that the excitation of the elementary particle in a constant electric field might be due to pair creation by electric field\cite{Nikishov}. Another problem is the time of formation of the thermal bath. If an electron appears in the electric field, as acceleration field, instantly what it will see? According to Ginzburg and Frolov\cite{Ginzburg} the electron will see the thermal bath only after a time far longer than $\tau= c/w$. What kind of process results in such thermalization has yet to be explained. Moreover, for photons the Unruh effect contains no Plank constant. According to 'conformity principle' for radiation, saying that any quantum radiation equals to classical one when $\hbar \rightarrow 0$, there should be a classical analog of such radiation. Mechanism of classical analog of Unruh radiation is difficult to determine.
 
 Nevertheless, there is a positive point that makes the study of the Unruh radiation interesting. The Unruh radiation has a strong analogy with Hawking radiation\cite{Hawking}. Whether the acceleration field is equivalent to the gravitational field is not verified, but the Unruh radiation might give us the answer to that is true or not.

 An unique complex SACLA consisting of 10 keV/30 fs XFEL systems \cite{SACLA} and coming $\approx$1 PW/25 fs Ti-Sph femtosecond laser with maximum intensity $I =10^{22}$ W/cm$^2$ {($a_0\approx 100$)} {, which exceeds the similar parameter of J-KAREN with peak intensity of $I =10^{22}$ W/cm$^2$ \cite{J-KAREN} which already exists,} allows developing a new, more adequate approach to explore the Unruh effect.

 One of the problems in detecting the Unruh effect is the necessity of emitters and detectors in the same Rindler reference frame. For this purpose, it is better to explore the photon scattering rather than photon emission. 

		\section{Thomson scattering from accelerated electrons}

 According to Unruh the emission probability of an electron in Rindler space is given by following equation\cite{Unruh1}:
	\begin{equation}
	P_j = {\displaystyle \sum_{p}}{ \left| \int_0^T\epsilon d \left ( \frac{\tau}{T} \right)  \int_V \sqrt{-g} d^3 x {\psi_j}^{\ast} \bra{p} A \ket{0}_M \psi_0  \right|}^2,
	\end{equation}
where $\psi$ is an electron wave function in Rindler space, $A$ is the field operator, $\bra{p}$ is the field wave function, $\epsilon$ is a coupling constant, $M$ means the vacuum in Minkowski space. This equation results in Rindler radiation. In the case of photon scattering one can extend the Unruh logic to a photon matrix element: $\bra{0}a_2 A_{\mu} (x)A_{\beta} (x^{\prime}) {a_1}^{\dagger} \ket{0}_M$ (see Ref.\cite{Landau2}, $a, a^{\dagger}$ are the known operators). It is clear that the result of Fulling-Rindler mode may occur only as a broadening of scattering massless photon similar to the Doppler broadening with the temperature $T_w=\hbar w/2\pi k_B c$.

 There was a proposal of using Thomson scattering as a detector for the Unruh effect\cite{Crowley} exploiting acceleration of electron quiver motion in a laser pulse, $ w = c \omega_L a_0$ with zero $v \times B$ force. To make the story, 3 different laser pulses were considered. 1st optical laser pulse should excite a gas-jet to make plasma. 2nd optical laser pulse makes acceleration for electrons in the plasma, and 3rd an XFEL pulse coaxial with the 2nd pulse probes accelerated electrons. It was stated that there is the possibility to detect the Unruh effect by obtained spectral broadening. 

 Expected spectrum is given by\cite{Crowley}
	\begin{equation}
	S({\bf k}, \omega) = \sqrt{ \frac{1}{\pi (1 - \cos \theta) \Omega^2}} \exp \left[ - \frac{1}{\pi (1 - cos \theta) \Omega^2} \omega^2 \right],
	\end{equation}
 where $\Omega = \omega_i v_{th} / c$, $\omega_i$ is the photon energy of probing laser, $\theta$ is scattering angle, $v_{th} = \sqrt{2k_B T_w  / m}$ is thermal velocity, $k_B$ is Boltzmann constant, $T_w = T + \lambda m {w^2} / { 2k_B} q^2 c^2$ is the temperature of electrons in thermal equilibrium, which differs from the Unruh temperature, $T$ is temperature of the plasma, $m$ is electron mass, $c$ is speed of light, $\omega$ is the change of photon energy of probing laser after scattering, $q$ is a change of wavenumber after scattering and $\lambda$ is metric parameter{, which determines the properties of space time geometry (metric) \cite{Crowley}}. About 40 photons are expected in a shot of XFEL. However, in this scheme, it is difficult to distinguish whether the expected spectrum broadening occurs due to the Unruh effect or due to the extra Doppler effect. {Because of} a finite length of probing pulses, electrons are further accelerated and evacuated by the ponederomotive force to vicinity of lower laser field as well as a strong, complicated longitudinal field make this scheme difficult in realization. We consider two schemes that can eliminate extra Doppler shifts and extract the Unruh effect using plasma and intense laser.

 		\section{Thomson scattering from a laser pulse wake wave}

Electrons in plasma wave exited by a laser pulse in underdense plasma are good candidates for probing the Unruh effect via Thomson scattering. It is well known that a laser pulse propagating in underdense plasma creates a trail of plasma wave in its wake  \cite{Bulanov1}. This wave has no group velocity that means plasma electrons in the wave oscillate with plasma frequency, $\omega_p = \sqrt{4 \pi Z N_p e^2 / m}$, where $N_p$ is the plasma density, $Z$ is the ion charge, and with a limited displacement. Such plasma perturbation propagates with phase velocity $v_p=c\sqrt{1-4 \pi ZN_p/a_0N_{cr}}$ where $N_{cr}$ is the critical density for laser pulse with frequency $\omega_L$ and $a_0$ is its normalized intensity.  In contrast to the electric field of a laser pulse, maximal electric field in a wake, and corresponding acceleration, is weaker by factor $\omega_p / \omega_0$ \cite{Esarey2}: $w = eE_p/m = c \omega_p a_0  $ [cm/s$^2$]. Moreover, for large $a_0$ this field strength can be achieved only in the first bucket of plasma wave trail due to the relativistic wave-breaking \cite{Bulanov2}. The maximal acceleration in regular wave buckets therefore is even smaller and is determined by the wave breaking limit field \cite{Esarey2}: $w = eE_{WB}/m = c \omega_p \sqrt{ 2 a_0}  $ [cm/s$^2$]. However, the number of electrons with identical velocity and moving at identical acceleration in these waves is much larger than that in the case of quiver motion \cite{Crowley}. This allows essential increasing of efficiency of Thomson scattering for detection of the Unruh effect. The problem of ponderomotive acceleration also vanishes. In the reference frame moving with the constant group velocity of wave this scheme provide an electron layer with $v=0$ undergoing constant acceleration $w$.

There are two parameters resulting in value of electron acceleration, the laser field strength, $a_0$, and the plasma density, $N_e$. 
However, one can consider three different cases of plasma wave for the Unruh effect detection: (i) a regular wake with minimal possible acceleration, $w = eE_{WB}/m$, and a large number of identical accelerated structures; (ii) wave breaking limit with far higher acceleration  $w  = c \omega_p a_0  $, but with a single regular accelerated electron group existing for a short time, (iii) relativistically transparent overdense plasma with immense number of accelerated electrons and maximal acceleration.

\begin{figure}[t]
\begin{center}
\includegraphics[width=\hsize]{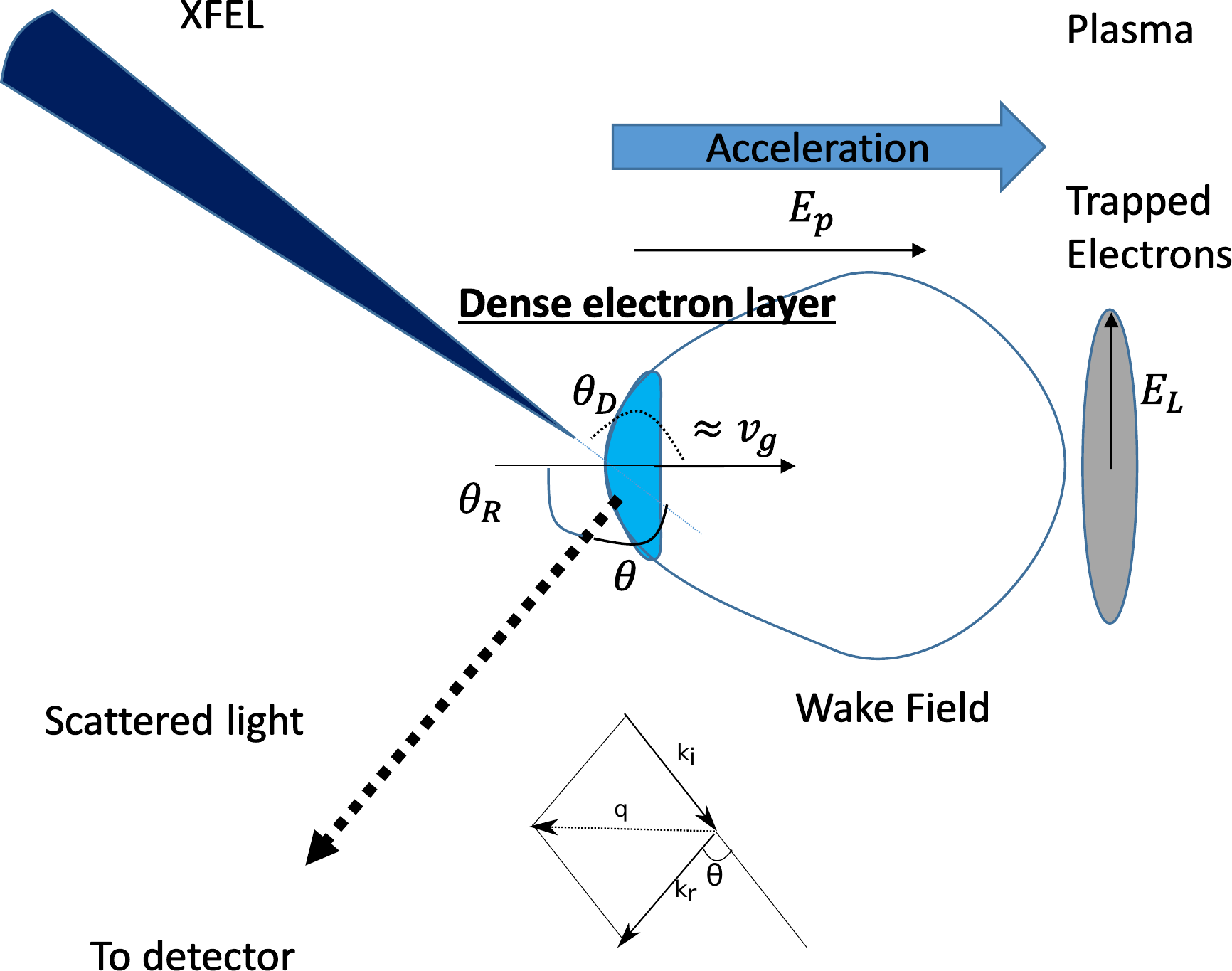}
\end{center}
\caption{}
\label{fig:fig1}
\end{figure}

 Being based on this, we propose the concept for observation of the Unruh effect by the electron layer of the plasma wave {with $k$ vector diagram \cite{Glenzer}} in Fig.\ref{fig:fig1}. The layer is produced in the plasma and accelerated by wake field. Probing laser is incident from its back and scattered light is detected also backwardly. {We use plasma wave similar to flying mirror \cite{Bulanov2} exploiting incident light from opposite direction to decrease intensity of laser light.} Akhiezer and Polovin\cite{Akhiezer} showed that electrons in plasma wave have acceleration $w(v)$ depending on its velocity. Upon choosing reasonable incident and detection angles, the Doppler shift of electron motion is suppressed and only spectral broadening due to the Unruh effect could be extracted.  The Doppler shift would be occurred when we observe the spectrum in the laboratory reference frame. That is shown as \cite{Bulanov2}
	\begin{equation}
	\varepsilon^{\prime} = \varepsilon_0 \frac{1 + 2 \beta \cos \theta_D + \beta^2}{1 - \beta^2} ,
	\end{equation}
where $\varepsilon_0$ and $\varepsilon^{\prime}$ are the photon energy of light before and after scattering respectively. $\beta$ is denoted as $\beta = v^2 / c^2$  where $v$ and $c$ are speed of electron layer and light respectively. $\theta_D$ is the angle between velocity of electron layer and incident light. And the reflection angle $\theta_R$ is related to the incident angle {shown in $k$ vector diagram of Fig.\ref{fig:fig1}} as
	\begin{equation}
	\sin \theta_R = \frac{\varepsilon_0}{\varepsilon^{\prime}} \sin\theta_D =\frac{(1 - \beta^2)\sin\theta_D}{1 + 2 \beta \cos \theta_D + \beta^2},
	\end{equation}
 
 We can find the angles that the scattering frequency should not be changed due to the Doppler shift from the above equations. This gives $\theta_D = 174^{\circ}$ and scattering angle $\theta_R = 5.4^{\circ}$. These angles we use for estimation of spectra of scattered x-rays.  When photon energy of probing laser is set to be 1 keV, expected spectral broadening is expected to be $\Delta\omega/\omega_i \approx 1.4 \%$ at $a_0 \approx 100 $, $N_e \approx 10^{20}$ cm$^{-3} $ for regular wake, $\Delta\omega/\omega_i \approx 3 \% $ at $a_0 \approx 100 $, $N_e \approx 10^{20}$ cm$^{-3} $ for first bucket. $\omega_p$ raises up to $\omega_L$ with increase of density. 

 To estimate efficiency of scattering we present the electron density in the form\cite{Akhiezer}
	\begin{equation}
	N_e \approx \eta(a_0)N_p,
	\end{equation}
where $\eta(a_0)$ is the compression coefficient depending on the laser intensity. We also define a parameter $\delta = w_{x-ray}/w_{L}$ as the uniformity parameter, which represents the flatness of the electron layer, where $w_{x-ray}$ and $w_{L}$ are the waists of x-rays and driving laser pulse respectively. {We solved the conventional transport equation $dN_{sc}/dx=N_e (d\sigma / d\Omega)N_{in}\delta$ assuming $x=L$ is too small, where $N_{in}$ is the number of incident photons. } Finally one can get $N_{sc}$ per 1 shot as
	\begin{equation}
	N_{sc}=N_e \frac{d\sigma}{d\Omega}N_{in}L\delta,
	\label{eq:numbe-sc}
	\end{equation}
 where $ d\sigma / d\Omega= {r_e}^2 (1 + \cos^2 \theta) / 2$ is the Thomson scattering cross section  with $r_e$ is the classical electron radius and $\theta$ is a scattering angles{, and $L$ is thickness of the layer. We use electrons in the plasma wave with high density. The wave moves with high speed as well as the x-rays. Therefore, the efficient length $L'$ becomes longer. For example in the case of incidence with zero angle the length will be $L'=Lc/(c-v)$ where $v$ is the phase velocity of the plasma wave. During this propagation there will be scattering. If we assume that the wave does not change much in this distance the number of photon drastically increases, at least 10-100 times. In the normal case the angle is not zero and efficient length is shorter. However, the number of photon increases.}

\begin{figure}[t]
\begin{center}
\includegraphics[width=\hsize]{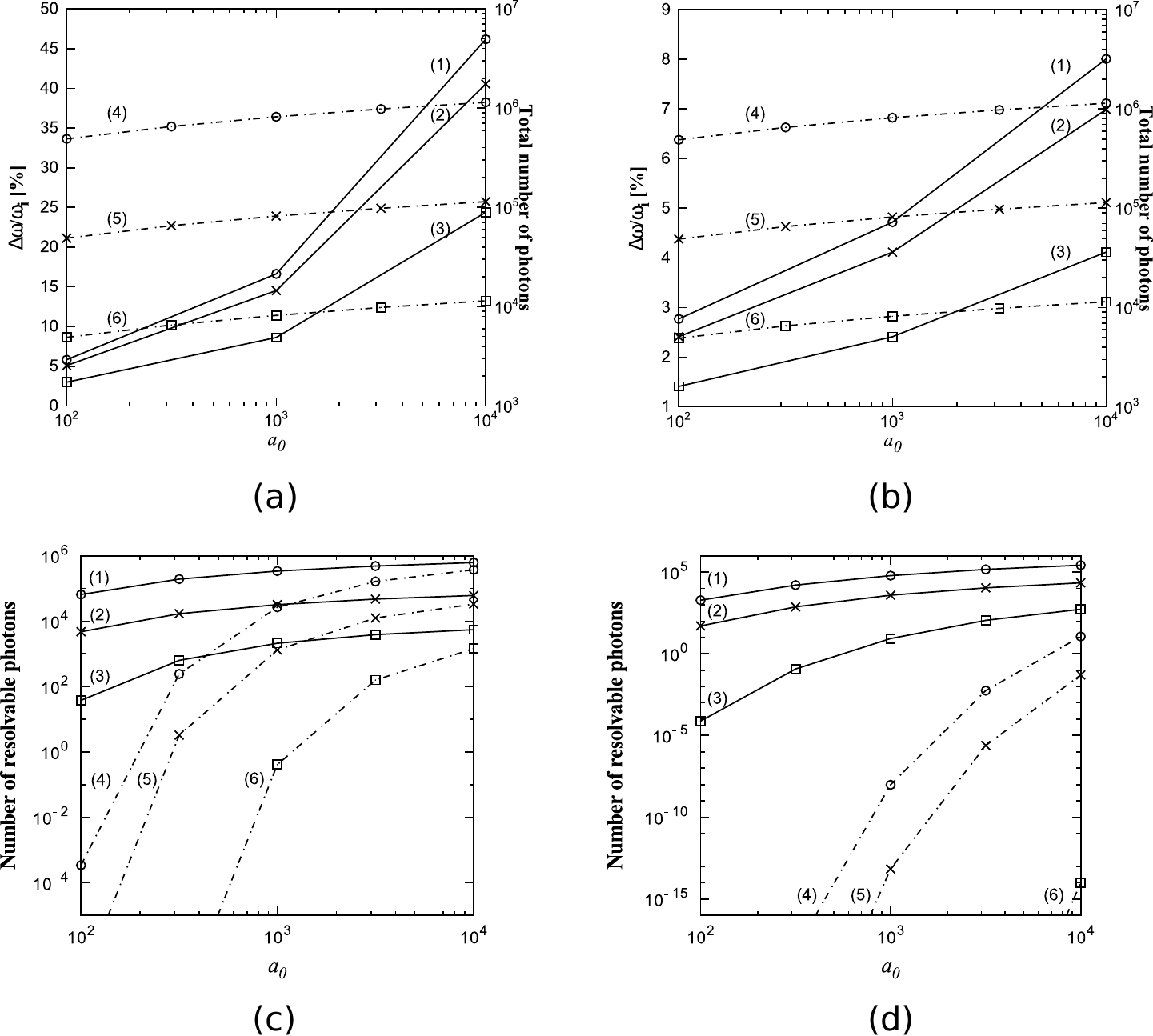}
\end{center}
\caption{}
\label{fig:fig2}
\end{figure}

Results of calculations for spectra and total number of scattered photons and its efficiency are presented in Fig.\ref{fig:fig2} for the maximal and minimal accelerations. The number of incident x-rays is $N_{in} = 10^{12}$, waists of x-rays $w_{x-ray}=7.5$ $\mu$ m and laser pulse waist $w_{L}=15$ $\mu$m{, and thickness $L\approx 1$ $\mu$m}. Efficiency is the number of resolvable photons in the total number of photons,  meaning that how many photons can be detectable for the Unruh radiation.  It is determined by the integral of spectra from resolvable point (1 \%, 5 \% in the case) to infinity.
 
\begin{figure}[t]
\begin{center}
\includegraphics[width=\hsize]{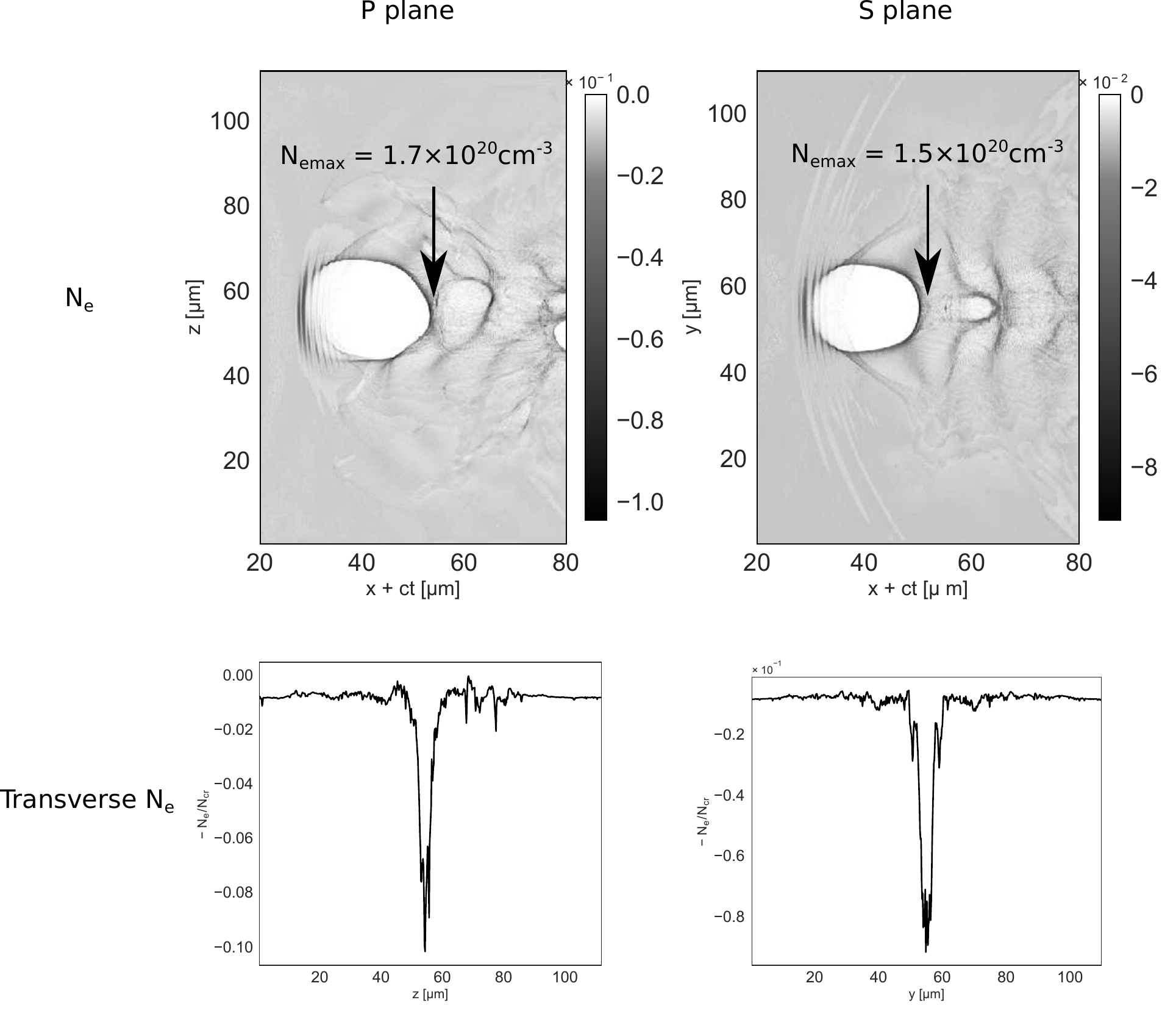}
\end{center}
\caption{}
\label{fig:pic}
\end{figure}

 This number of photons is enough to get reasonable results by 1 shot probing. The case of relativistically under-dense plasma has been included in this calculation. The compression efficiency was calculated using 3D PIC FPLASER3D code\cite{Zhidkov}. One can see that the efficiency of scattering is high enough for verifiable detection of the Unruh effect in both cases. The number of scattering photons exceeds $10^{3}$ even for $a_0 = 100$ and the plasma density $N_p = 10^{20}$ cm$^{-3}$. {We show  in Fig.\ref{fig:pic} that shows the results of 3D PIC simulation when intense laser pulse with $a_0 = 100$ and wavelength of 1 $\mu$m irradiated to underdenseplasma (hydrogen gas in this case) with initial electron density of $10^{19}$ cm$^{-3}$. The critical density is approximately $1.7\times 10^{21}$ cm$^{-3}$. We see the electron density of the layer exceeds $10^{20}$ cm$^{-3}$ because of the compression.}  Expectedly the broadening of the scattered x-rays owing to the Unruh effect is larger for the maximal acceleration. Nevertheless even for $a_0 = 100$ and the plasma density $N_p = 10^{20}$ cm$^{-3}$ it is about $1\%$ that could be enough for detection depending on the broadening of XFEL pulses. 

 We note that this scheme is free from recoil problem. For co-propagating x-rays and laser pulse recoil effect is small $\hbar \omega / 2 \gamma m c^2 \approx 10^{-4}$ for $\gamma = 10$ and 1 keV x-ray. The use of the first bucket of wake field would provide us with the maximal acceleration and give parameters of scattered x-rays good enough for detection of the Unruh effect. However, due to relativistic wave breaking a regular electron layer exists only though for a short time but quite detectable even in the case of high density plasma\cite{Kulagan}. This requires a tight synchronization of laser pulse and XFEL pulse. Full optical XFEL system\cite{ImPACT}can provide such synchronization down to a few fs level.
 
 		\section{Thomson scattering from electron bunches accelerated by laser wake field}

\begin{figure}[t]
\begin{center}
\includegraphics[width=\hsize]{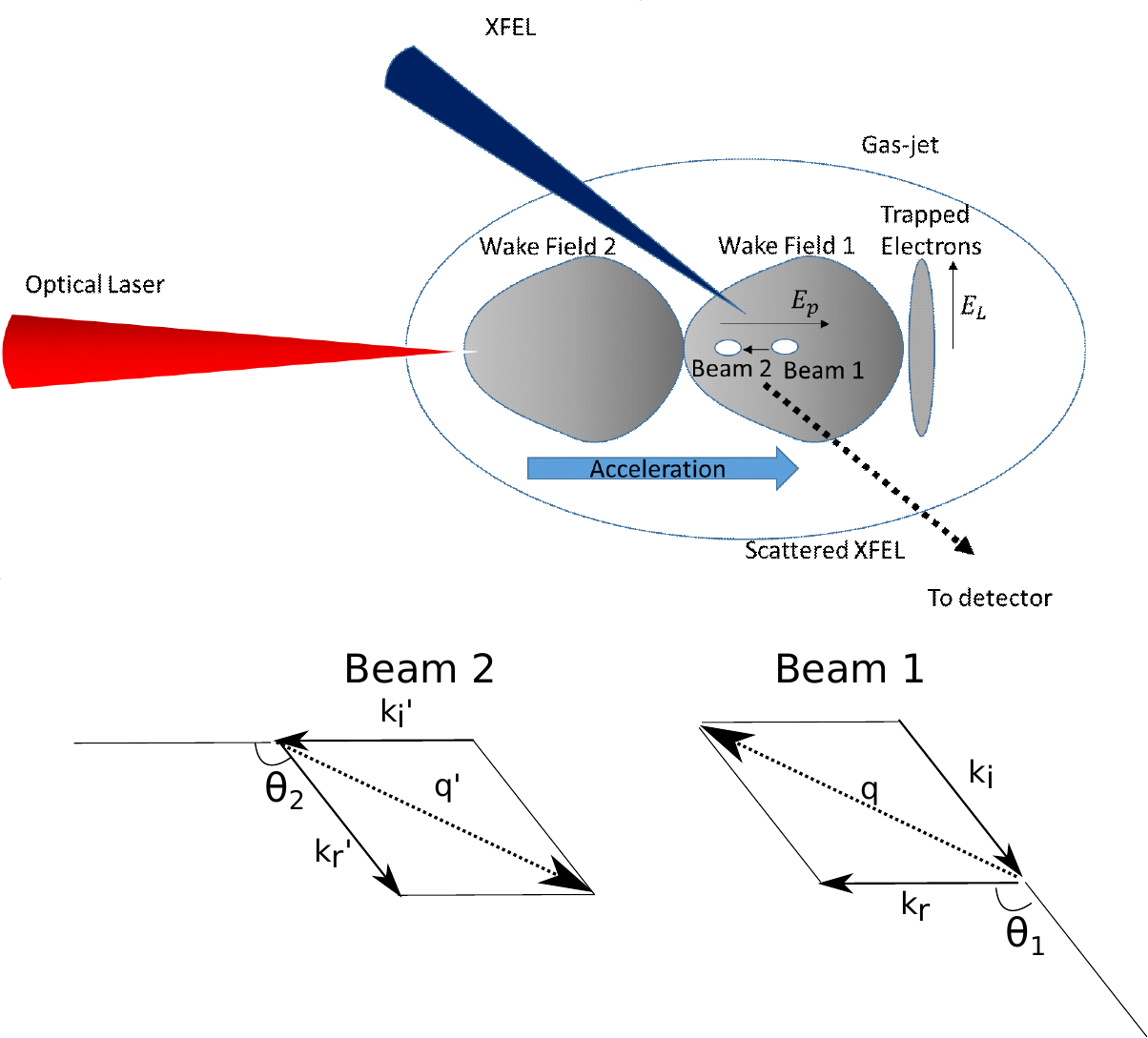}
\end{center}
\caption{}
\label{fig:fig3}
\end{figure}

 The use of Thomson scattering from a laser wake wave demonstrates its serious prospect for detection for the Unruh effect. One contradiction, however, is based on the fact that measurements should include not only source of radiation but also detector in the same reference frame. Here, we propose such a system consisting of a source and detector also with laser wake field acceleration (LWFA) and XFELs. Proposed experimental setup is shown in Fig.\ref{fig:fig3}. Pumping optical laser irradiates a helium gas-jet to create laser wake field. Due to multiple electron self-injection several beams, for example, electron beam 1 and electron beam 2 shown in Fig.\ref{fig:fig3}, are created in an accelerating field $E_P$ and can be used as a source and a detector respectively. Probing XFEL (attosecond harmonics with 1 eV spectrum width) is incident to an accelerated source from its back. Scattered light from the detector is probed and the spectrum is to be obtained.

\begin{figure}[t]
\begin{center}
\includegraphics[width=\hsize]{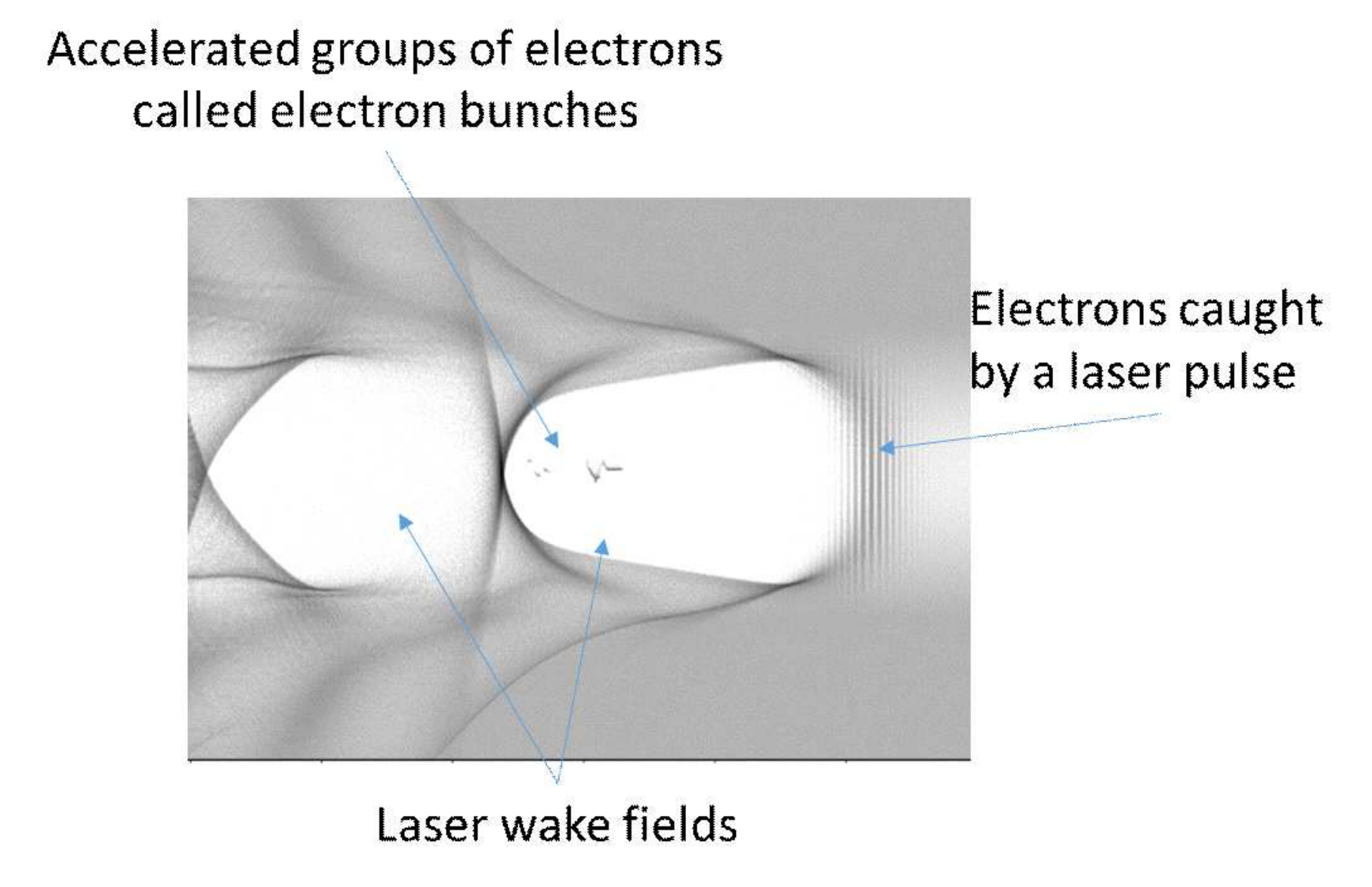}
\end{center}
\caption{}
\label{fig:fig4}
\end{figure}

 Recently, electron acceleration technology has been developing. LWFA can accelerate electrons with an acceleration gradient of $\approx 100$ GV/cm. A laser pulse leaves its wake field blowing electrons away in plasma\cite{Zhidkov}. Acceleration field is generated by sparse and dense electrons. Some of blown electrons come back to acceleration field by Coulomb force and built groups of electrons, which is called bunch, are uniformly accelerated with the same high gradient acceleration and velocity. {Uniform acceleration means that all electrons are under the same acceleration force in electric field with its gradient equaling to zero and  in ideal case lower gamma is better for such purpose because of relativistic effects with higher gamma. What we need is large acceleration not energy.} Thus, LWFA is adequate to verify the Unruh effect. Fig.\ref{fig:fig4} shows the typical result of 2D PIC simulation for laser wake field acceleration. Due to multiple electron self-injection\cite{Zhidkov} or an external injection, the number of bunches may be varied.
 
 Accelerating electrical field is the wake field with its strength given by
	\begin{equation}
	w = \frac{e}{m} E_p = c \omega_p a_0 \ \ [\text{cm/s}^2],
	\end{equation}

\begin{figure}[t]
\begin{center}
\includegraphics[width=\hsize]{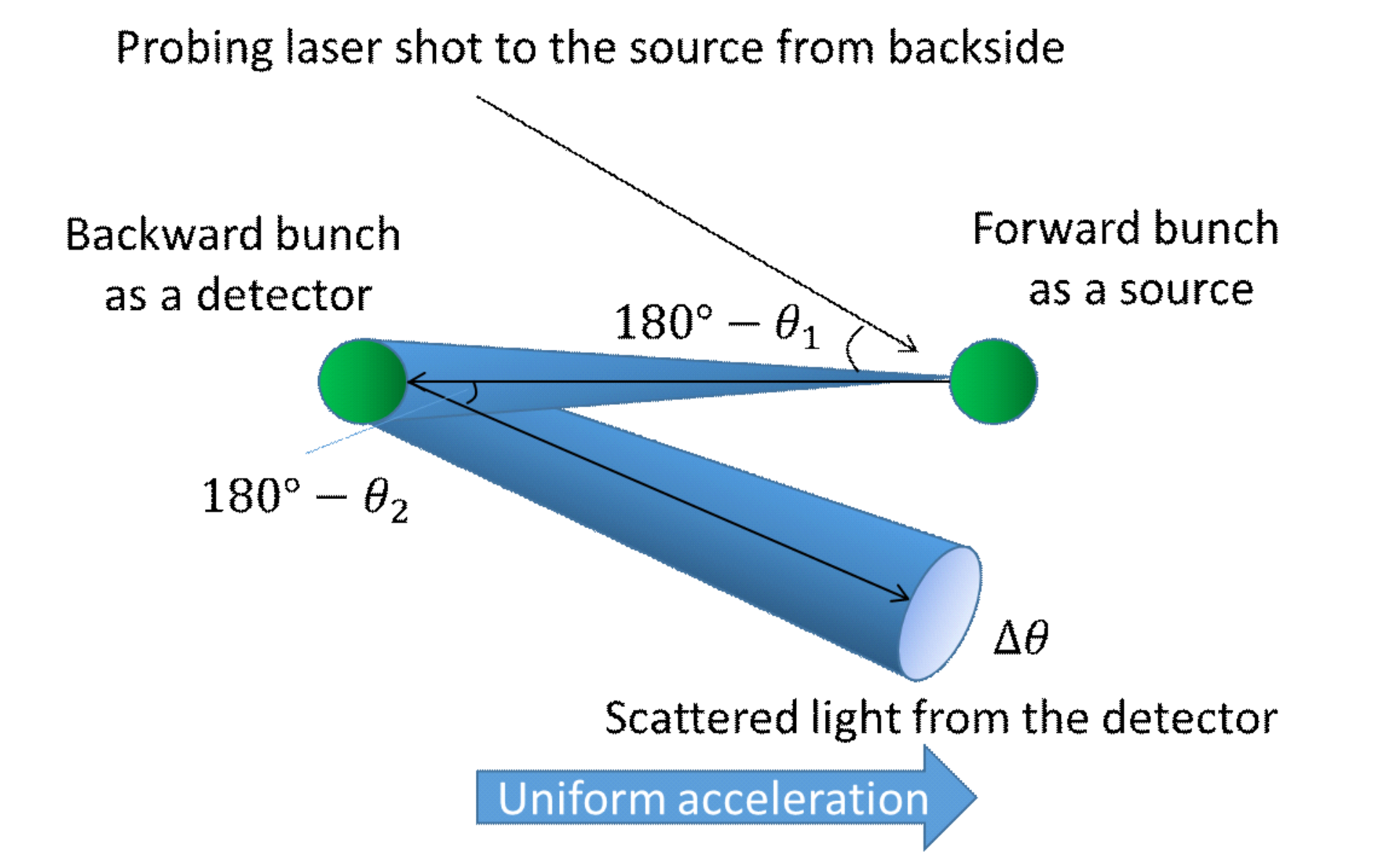}
\end{center}
\caption{}
\label{fig:fig5}
\end{figure}

Again, we consider Thomson scattering from bunches to observe the Unruh effect. Figure.\ref{fig:fig5} shows the experimental process {with $k$ vector diagram \cite{Glenzer}}. When we consider Thomson scattering from 1 bunch, scattered spectrum broadens due to velocity of bunches. To eliminate this effect, we take a pair of 2 bunches as source and detector and consider the 2 times Thomson scattering. We use a forward bunch as a source and backward bunch as a detector. Electron beams 1 and 2 are accelerated by different acceleration $w_1$ and $w_2$ that have almost the same value. We could put the source and detector in almost the same accelerating frame. Probing laser is shot from the back side of the source. The source scatters photons to the detector. We observe the spectrum of scattered photons by the detector. Frequency of scattered light by source becomes downshifted due to Doppler effect. And frequency of scattered light by detector becomes upshifted due to Doppler effect. So, observed spectrum of detector should be the same as initial probing pulse ordinary. However, if the Unruh effect exists, accelerated bunches feel the Unruh temperature and the spectrum of scattered light becomes broader. We could extract the Unruh effect felt by each electron beams.

Scattering occurs twice by the source and the detector. The frequency of the probing laser becomes shifted in the first scattering by the source. The detector scatters the shifted laser light from the source returning its initial frequency. So, the expected spectrum from the detector, observed in the laboratory frame, should be obtained by convolution of $S({\bf k},\nu)$
	\begin{equation}
	\begin{split}
	S({\bf k},\omega) & \approx   \int_{-\infty}^{\infty} S({\bf k},\nu) S({\bf k},\omega - \nu) d\nu \\
&= \sqrt{\frac{1}{\pi F(\theta)}} \exp [- G(\theta) \omega^2],
	\end{split}
	\end{equation}
	\begin{equation}	
	F(\theta) = (1-\cos \theta_2) {\Omega_2}^2+(1-\cos \theta_1) {\Omega_1}^2,
	\end{equation}
	\begin{equation}
	G(\theta) = \frac{1}{(1-\cos \theta_1) {\Omega_1}^2}- \cfrac{1}{{(1-\cos\theta_1)}^2  {\Omega_1}^4 \left[ \cfrac{1}{(1-\cos \theta_1) {\Omega_1}^2} + \cfrac{1}{(1-\cos \theta_2) {\Omega_2}^2} \right]},
	\end{equation}
	\begin{equation}
	\Omega_1 = \frac{\omega_i v_{th1}}{c}, \ \ \Omega_2 = \frac{\omega_i v_{th2}}{c},
	\end{equation}
where  $\theta_1$ and $\theta_2$ is scattering angle of source and detector, $v_{th1}=\sqrt{2k_B T_{w1} / m}$ and $v_{th2}=\sqrt{2k_B T_{w2} / {m}}$, $T_{w1} = \hbar {w_1} / 2\pi k_B c$ and  $T_{w2} = \hbar {w_2} / 2\pi k_B c$ are the Unruh temperature of beam 1 and 2, respectively.

 We have to set adequate scattering angles to eliminate the normal Doppler effect of scattered light. That means the scattering angles that we observed spectrum becomes monochromatic. Such scattering angles satisfy the equation shown below.
	\begin{equation}
	\cos(\pi - \theta_2) = \frac{1}{\beta} \left[ 1 - \frac{1-\beta^2}{1 - \cos\theta_1} \right],
	\end{equation}
where $\beta = v^2 / c^2$ and $v$ is the velocity of bunches. When $v=0.9c$, a set of angles satisfying the equation is $\theta_1=165^\circ$, $\theta_2=172.4^\circ$. {The angles are described in $k$ vector diagram in Fig.\ref{fig:fig3}}
 
 The spectral width of the expected spectrum is a critical parameter. Broadening due to the Unruh effect must exceed both a bandwidth of XFEL pulses and a thermal-like Doppler shift owing the energy spread of electrons in the beams. Broadening of XFEL can be solved upon use of harmonics from laser plasma\cite{Pirozhkov} while the problem of low beam energy spread is far from solution.

 The number of scattered photon per 1 shot from a single pair of source and detector can be estimated only by use of results of 3D PIC simulation shown in Fig.\ref{fig:fig4}. Density of photons from probing XFEL can be evaluated from a simple equation as $N_{in} = I_p  \pi {(w_U/ 2)}^2 \tau / E_{ph}$ , where $I_p$, $w_U$, $\tau$ and $E_{ph}$ are laser intensity, beam waist, pulse duration, and photon energy respectively. We denote the number of scattered photons from the source as $N_{sc1}$, the number of scattered photons by the detector as $N_{sc2}$, and the electron density of bunch as $N_e$. Since the interaction length $L$ is equal to the size of the bunch, $N_{sc1}$ is given by $N_{sc1} = N_e  (d\sigma / d\Omega)_1 N_{in} L$. Then, $N_{sc2}$ becomes
	\begin{equation}
	\begin{split}
	N_{sc2} &= N_e \left( \frac{d\sigma}{d\Omega} \right)_2  N_{sc1} L \\
&= {N_e}^2 \left( \frac{d\sigma}{d\Omega} \right)_2 \left( \frac{d\sigma}{d\Omega} \right)_1 L^2 N_{in}, \\
	\label{eq:beams}
	\end{split}
	\end{equation}
{in ideal case} where $ (d\sigma / d\Omega)_i = {r_e}^2 (1 + \cos^2 \theta_i) / 2$ is the Thomson scattering cross section of each scattering, with $r_e$ is the classical electron radius and $\theta_i$ is a scattering angles. {In this case the efficient length as discussed in the previous section could be considered. Electron beams move with high speed as well as the x-rays. The efficient length $L'$ becomes longer. As we showed, in the case of incidence with zero angle the length will be $L'=Lc/(c-v)$ where $v$ is the phase velocity of the electron beams.} Upon setting $N_{in} = 10^{12}$  from the equation, $N_e = 10^{20}$ cm$^{-3}$ and $L = 10$ $\mu$m from the results of 3D PIC simulation,  one can get $N_{sc2} \approx 5 \times 10^{-5}$ in 1 shot. This number means that we can get $10^{-5}$ photons of the spectrum per 1 shot. Since several pairs of bunches can be generated, the number of photons can be increased proportionally.
 
 Let us estimate the number of pairs we can use when using SACLA's XFEL that has 10 keV and $\approx$ 80 $\mu$m beam size. Size of the bunch is $\approx$ 2 $\mu$m, 2 bunches are used as a pair. A wavelength of light scattered by a source becomes $\approx$ 0.8 $\mu$m. So, a pair needs $\approx$ 4.8 $\mu$m. The beam size of XFEL is $\approx$ 80 $\mu$m. Therefore, $\approx$ 30 pairs of bunches can be used in 1 shot probing. When we have 30 pairs of bunches, we could get $N_{sc2} \approx 1.5\times 10^{-3}$.
 We can obtain sufficient spectrum by probing many times. For example, if we use the laser in 10 Hz repetition rate mode, the expected number of photons is $\approx 5$ per hour. It seems to be a sufficient number of photons to resolve the scattered spectrum.

{In this case, the charge of electron beams is smaller and of order nC-pC.  Because of low charge of the beams, this experiment would be sensitive to scattering compared to using electrons in plasma wave proposed in previous section. In previous section, the number of electrons in plasma wave is large enough to obtain spectrum and charge is not so important. In using electron beams case, electron charge is important. Small charge electron beams would not scatter the enough number of photons to obtain spectrum while eq.(\ref{eq:beams}) is assumed that second scattering scatters all photons from first scattering.}

\begin{figure}[t]
\begin{center}
\includegraphics[width=\hsize]{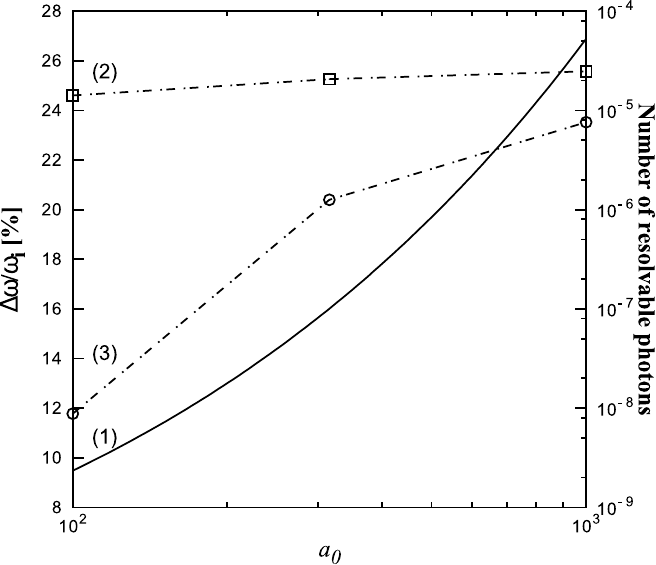}
\end{center}
\caption{}
\label{fig:fig6}
\end{figure}

 To verify the existence of the Unruh radiation, scattered spectrum should have certain width that relates to the acceleration. Fig.\ref{fig:fig6} shows $a_0$ dependency of spectrum width. As seen from this graph, higher acceleration is necessary to observe broader spectrum.

		\section{Conclusions}

 We have proposed designs for experimental observation effects by spectral shift of Thomson scattering from two systems consisting of highly accelerated electrons. First system is consisted of electrons in the plasma wave excited by an intense laser. Reasonable photon number of Thomson scattering allows us to obtain the spectrum. Choosing certain incident and detection angles enables us to distinguish the Unruh effect from the normal Doppler shift. Second system is consisted of source-detector moving with a constant acceleration. In this case, the laser wake field with the multiple electron self-injection has been considered. Probing the existence of the Unruh effect explores whether acceleration and gravitation are equivalent. Although there are several problems in the existence of the Unruh effect, it should be proved in the experiment. Two proposals enable us to distinguish spectral shift of the Unruh effect from that of the normal Doppler shift. Second proposed experiment has a source and a detector in the reference frame with constant acceleration and is suitable for detecting the effect occurring in the reference frame. The proposed experiment can prove whether the Unruh effect exists or not. 

 New proposals can be realized at SACLA \cite{SACLA} in near future. {Lasers with $a_0\approx100$ is coming to SACLA exceeding already achieved laser parameter of J-KAREN \cite{J-KAREN}.} For now, the maximum intensity of optical laser reached $a_0\approx100$ experimentally {in the world} and is developing to reach higher intensity. In the case of plasma wave the firm detection of the Unruh effect can be performed soon with $a_0 \approx 100-200$. Therefore, this proposal could be executed by a combination of XFEL and LWFA with intense optical laser theoretically. The technologies of LWFA and X-ray harmonics are under development but are not  yet practical.

The systems we proposed are considered to leave only practical problems, such as attosecond harmonics, Laser Wake Field Acceleration, synchronization between PW laser and XFEL and the facility consisting of all these technologies. IMPACT project is running to construct the facility that makes it possible in SACLA. Fundamental problems seem to be solved to verify the Unruh effect. We can expect to execute the proposed experiment at SACLA facility with the PW class laser system and the existing XFEL.
 
 		\section{Acknowledgement}

 This work was partially supported by IMPACT project.

 \newpage
{\bf Figure Captions}\\ \\
Fig. 1. Proposed experimental setup using accelerated electrons of plasma wave {with $k$ vector diagram. $k_i$ and $k_r$ is wave vector of incident and scattered light, respectively and $q$ is change of wave vector.} The layer is produced in the plasma and accelerated by wake field. Probing laser is incident from its back and scattered light is detected also backwardly.Upon choosing reasonable incident and detection angles, the Doppler shift of electron motion is suppressed and only spectral broadening due to the Unruh effect could be extracted.\\ \\
Fig. 2. Dependency of spectrum width and the total number of photons on the laser pulse field for (a) scattering from the first bucket at maximal acceleration and (b) for the wave breaking limit. Solid lines show $\Delta \omega / \omega_i$ for $N_e =$ (1)$10^{22}$, (2)$10^{21}$ (3)$10^{20}$ cm$^{-3}$. Dashed lines show the total number of photons for $N_e =$ (4)$10^{22}$, (5)$10^{21}$ (6)$10^{20}$ cm$^{-3}$. The number of resolvable photons for (c)the first bucket and (d)the wave breaking limit; the photon energy of probing laser is $\omega_i=1$ keV. Solid lines show the number of resolvable photons with 1 \% resolution for $N_e =$ (1)$10^{22}$, (2)$10^{21}$ (3)$10^{20}$ cm$^{-3}$. Dashed lines show the number of resolvable photons with 5 \% resolution for $N_e =$ (4)$10^{22}$, (5)$10^{21}$ (6)$10^{20}$ cm$^{-3}$.\\ \\
Fig. 3. {Results of 3D PIC simulation when intense laser pulse with $a_0 = 100$ and wavelength of 1 $\mu$m irradiated to underdense plasma (hydrogen gas in this case) with electron density of $N_p = 10^{20}$ cm$^{-3}$. The critical density is approximately $1.7\times 10^{21}$ cm$^{-3}$. We see the electron density of the layer exceeds $10^{20}$ cm$^{-3}$ around 
$x+ct=55$ [$\mu$m] which is focal position of the laser pulse because of the compression.}\\ \\
Fig. 4. Proposed experimental setup with a source and a detector. {$k$ vector diagram is shown. $k_i$ and $k_i$ is wave vector of incident and scattered light, respectively and $q$ is change of wave vector of first scattering and with prime is of second scattering.} $E_L$ is electric field of optical laser pulse. $E_P$ is accelerating field of wake field. Beams 1 and 2 are group of electrons used as a source and a detector respectively. Pumping optical laser ( $\geq10^{22}$ W/cm$^{-2}$) is shot in Helium gas-jet to create Laser Wake Field. Probing XFEL is incident to an accelerated source from its back. Scattered light from the detector is probed and the spectrum is to be obtained.\\ \\
Fig. 5. 3D simulation for Laser Wake Field Acceleration. Electron density from 3D PIC simulation for LWFA done for the conditions stated below. The plasma channel has the maximum density of $N_{e max} = 4\times10^{18}$ cm$^{-3}$ and the minimal density of $N_{e max} = 1.2\times10^{18} $ cm$^{-3}$. The laser intensity is $I = 4 \times 10^{19}$ W/cm$^{-2}$, the focus diameter is $D = 18$ $ \mu$m and laser pulse duration is 30 fs. One can see multiple self-injection of bunches in the first bucket behind the laser pulse. Some bunches appear in the bucket repeatedly.\\ \\
Fig. 6. Proposed experimental process. Forward bunch could be used as source and backward bunch as a detector. Probing laser is irradiated from the back side of the source. Source scatters photons to the detector. We observe the spectrum from the detector. Electron bunches moving with the same acceleration.\\ \\
Fig. 7. Dependency of spectrum width and efficiency of number of photons on the laser pulse field. The graph shows the dependency of spectrum width and efficiency on the laser pulse field strength $a_0$ . $\omega_i$ is the photon energy of probing laser set to be 1 keV. Solid line (1) shows  $\Delta \omega / \omega_i$. Dashed lines show the number of resolvable photons with (2)1 \% (3)5 \% resolution. The spectral width means the strength of the the Unruh effect. To obtain broader spectral shift, the larger $a_0$ is required.\\ \\
 
\clearpage


\clearpage

\clearpage

\clearpage

\clearpage

\clearpage

\clearpage

\end{document}